\documentclass[english,conference]{IEEEtran}
\usepackage[T1]{fontenc}
\usepackage[latin9]{inputenc}
\usepackage{color}
\usepackage{babel}
\usepackage{amsmath}
\usepackage{amsthm}
\usepackage{amssymb}
\usepackage{graphicx}
\usepackage[unicode=true,pdfusetitle,
 bookmarks=true,bookmarksnumbered=false,bookmarksopen=false,
 breaklinks=false,pdfborder={0 0 0},pdfborderstyle={},backref=false,colorlinks=true]
 {hyperref}

\makeatletter
\theoremstyle{plain}

\theoremstyle{remark}

\theoremstyle{definition}

\theoremstyle{assumption}

\usepackage[ruled]{algorithm2e}
\newcommand{\nosemic}{\renewcommand{\@endalgocfline}{\relax}}
\newcommand{\dosemic}{\renewcommand{\@endalgocfline}{\algocf@endline}}% Reinstate semi-colon ;
\usepackage{cite}
\usepackage[paper=letterpaper,top=0.75in,bottom=0.75in,right=0.75in,left=0.75in]{geometry}
\usepackage{subfigure}
\usepackage{enumerate}
\newtheorem{assumption}{Assumption}

\makeatother

\newtheorem{remark}{Remark}

\newtheorem{theorem}{Theorem}

\providecommand{\definitionname}{Definition}
\providecommand{\remarkname}{Remark}
\providecommand{\theoremname}{Theorem}

\begin{document}
\newgeometry{top=1in,bottom=0.75in,right=0.75in,left=0.75in}
\IEEEoverridecommandlockouts 
\title{Adaptive Actor-Critic Based Optimal Regulation for Drift-Free Uncertain Nonlinear Systems}
\author{Ashwin P. Dani, Shubhendu Bhasin \thanks{A. P. Dani is with the Department of Electrical and Computer Engineering at University of Connecticut, Storrs, CT 06269. S. Bhasin is with the Indian Institute of Technology Delhi Email: ashwin.dani@uconn.edu, sbhasin@ee.iitd.ac.in}}
\maketitle

\begin{abstract}
	In this paper, a continuous-time adaptive actor-critic reinforcement learning (RL) controller is developed for drift-free uncertain nonlinear systems. Practical examples of such systems are image-based visual servoing (IBVS) and wheeled mobile robots (WMR), where the system dynamics includes a parametric uncertainty in the control effectiveness matrix with no drift term. The uncertainty in the input term poses a challenge for developing a continuous-time RL controller using existing methods. In this paper, an actor-critic or synchronous policy iteration (PI)-based RL controller is presented with a concurrent learning (CL)-based parameter update law for estimating the unknown parameters of the control effectiveness matrix. An infinite-horizon value function minimization objective is achieved by regulating the current states to the desired with near-optimal control efforts. The proposed controller guarantees closed-loop stability and simulation results validate the proposed theory using IBVS and WMR examples.
\end{abstract}

\section{Introduction}
Drift free dynamics are commonly found in robotics and other engineering applications. These are systems of the form $\dot{x} = g(x)u$. Some examples of such systems are image-based visual servo (IBVS) control, wheeled mobile robot (WMR), models of kinematic drift effects in space systems \cite{leonard1995motion} etc. Reinforcement learning (RL) has successfully provided a means to design optimal adaptive controllers for various classes of systems \cite{bertsekas1996neuro,jiang2017robust,sutton2018reinforcement,prokhorov1997adaptive}. For the drift-free systems, when there is uncertainty in the drift term, existing continuous-time RL solutions cannot be applied to design an RL policy. In this paper, an adaptive actor-critic method is developed for a class of drift-free nonlinear systems with uncertainty in the control effectiveness matrix.

RL learns the optimal policy that maximizes a long term reward. By interacting with the environment the decision maker gets evaluative feedback about its actions, which is used to improve the control policy \cite{kiumarsi2017optimal}. A popular class of iterative RL methods is adaptive dynamic programming (ADP), which was introduced by Werbos for discrete-time (DT) systems \cite{jiang2017robust,chen2008generalized,werbos1992approximate}, and implemented in actor-critic (AC) framework. Extension of RL algorithms to continuous time systems is achieved in \cite{doya2000reinforcement} by using Hamilton-Jacobi-Bellman (HJB) framework with known system dynamics, where a continuous-time version of the temporal difference error is employed. Several offline approaches for solving a generalized HJB equation are developed in \cite{beard1997galerkin, abu2005nearly}, using Galerkin's spectral approximation \cite{beard1997galerkin} and least-squares successive approximation solution \cite{abu2005nearly} to HJB, which is then used to compute the optimal control.

When the system dynamics are not completely known, among online approaches, an integral reinforcement learning (IRL) method is developed in \cite{vrabie2009neural,vrabie2009adaptive}, which requires only partial knowledge of system dynamics. The approach called policy iteration (PI) is developed based on AC structure, where the actor neural network (NN) is learned at a faster time scale than the critic NN. In \cite{vamvoudakis2010online} the IRL approach is extended to simultaneously learn both the actor and critic NNs, leading to a new method called synchronous PI. Further in \cite{bhasin2013novel}, an actor-critic-identifier (ACI) approach is presented, which in addition to actor and critic NNs, uses an identifier network to identify the unknown drift term in the dynamics. A model-based PI algorithm is developed in \cite{kamalapurkar2016model} for unknown drift dynamics where concurrent learning (CL)-based adaptive parameter update law is used to identify the drift part of the dynamics. In \cite{yang2021robust}, a robust actor-critic RL policy is developed for a class of nonlinear systems where certain parameterized unknown parts of the dynamics are estimated using adaptive update law. The above-mentioned methods require a complete knowledge of the input gain or control effectiveness matrix. 

The method in \cite{modares2013adaptive} identifies the complete nonlinear system dynamics using the experience replay technique and learns the actor-critic NN using PI method for completely unknown dynamics. For linear systems, a completely model-free RL method is developed in \cite{jiang2012computational}, which iteratively solves the algebraic Riccati equation using the online information of state and input. Using the IRL framework, an on-policy model-free Q learning approach is developed in \cite{vamvoudakis2017q} for linear systems. A data-driven approach to the actor-critic algorithm is developed in \cite{wang2015data}, where the system dynamics are identified using an NN, and the $g(x)$ term is recovered from the identifier before using it in the AC structure. In \cite{sahoo2016approximate} an identifier is used to identify the dynamics in the PI structure when the state and control input data are available at sampled time instances. A backstepping technique is used in \cite{li2019adaptive} to design a finite-time stabilizing controller with unknown dynamics parameterized using NN, which is then proved to be optimal with respect to a performance criterion. These methods are based on using an identifier structure to identify the unknown dynamics that are used in the AC structure, whereas an adaptive parameter update law using the CL method is designed for model parameter estimation in our paper.

The contribution of this paper is to design an adaptive actor-critic (AAC) RL algorithm for a class of uncertain drift-free nonlinear systems. The uncertainty in the control effectiveness matrix complicates the design of the PI RL algorithm for the drift-free systems. The unknown parameters of the system dynamics (input gain/control effectiveness) are modeled as a constant parameter for which a concurrent learning (CL)-based adaptive parameter update law is developed. The CL method uses a finite excitation condition which can be verified in real-time. The critic and actor NNs approximate the optimal value function and optimal control, respectively. The critic NN weight update law is derived based on the minimization of the Bellman error computed using optimal and approximate HJB equation. A least-squares weight update law is derived, which uses a persistency of excitation (PE) condition of the critic regressor. Similarly, a gradient-based NN weight update law is derived for actor NN based on the minimization of Bellman error. The parameter and actor-critic weights and the system model are learnt simultaneously as new state and control input data become available. Lyapunov stability analysis shows an exponential convergence of the state and parameter estimation errors to an ultimate bound, leading to uniformly ultimately bounded (UUB) stability. The proposed AAC policy is validated using two simulation examples, IBVS and WMR. In both simulation studies, the controller regulates the system state to its desired value. Compared to our prior work in \cite{dani2023reinforcement}, where an RL controller is designed for IBVS system, this paper derives the RL-policy for a generalized case of vector model parameter using vector parameter update law derived using CL \cite{chowdhary2013concurrent}.

\section{System Model and Control Objective}
\subsection{System Dynamics}
Consider the following system representing the evolution of the states as a function of the control input. The system model can be written in the following form
\begin{equation}
\dot{x}=g(x,\theta)u \label{eq:SystemModel}
\end{equation}
where $x(t) \in \mathbb{R}^{n}$ is the state, $u(t) \in \mathbb{R}^m$ is the control input, $\theta \in \mathbb{R}^{p}$ is an unknown parameter vector. The input gain matrix $g(x,\theta) \in \mathbb{R}^{n \times m}$ is expressed in parametric form as
%\begin{equation}
$\mathrm{vec}(g(x,\theta)) = Y(x)\theta$, %\label{eq:LinParam}
%\end{equation}
where $Y \in \mathbb{R}^{nm \times p}$ is a regressor matrix, $\mathrm{vec}:\mathbb{R}^{n\times m} \rightarrow \mathbb{R}^{nm}$ is a vectorization operator and $\mathrm{vec}^{-1}:\mathbb{R}^{nm} \rightarrow \mathbb{R}^{n \times m}$ is an inverse vectorization operator.
Using the parametric form of $g$, and properties of Kronecker product, the dynamics in (\ref{eq:SystemModel}) can be written as linear-in-parameter (LIP) form, given by
\begin{equation}
    \dot{x}=\mathcal{Y}\theta   \label{eq:SystemLIPForm}
\end{equation}
where $\mathcal{Y} = (u^T \otimes I_{n\times n})Y(x)$.

\subsection{Controller Objective}
The control objective is to regulate the current state to the desired state, $x_d\in \mathbb{R}^{n}$.
For the control design, the regulation error $\bar{x}(t) \in \mathbb{R}^{n}$ is defined as\begin{equation}
    \bar{x}(t) \triangleq x(t) - x_d
    \label{eq:regulation_error}
\end{equation}and the parameter estimation error $\tilde{{\theta}}(t)\in\mathbb{R}^p$ is defined as
\begin{equation}
\tilde{\theta}(t) \triangleq \theta-\hat{\theta}(t). \label{eq:thetaError}
\end{equation}
where $\hat{\theta}(t) \in \mathbb{R}^p$ is the parameter estimate. Since the optimal regulation objective is to bring the state $x(t)$ to a non-zero desired state $x_d$, the system model (\ref{eq:SystemModel}) is first written in terms of $\bar{x}(t)$
\begin{align}
    \dot{\bar{x}} = g(x,\theta)u 
    =g(\bar{x},x_d,\theta)u. \label{eq:ModSystem}
\end{align}
A continuous adaptive actor-critic controller is designed using the system in (\ref{eq:ModSystem}) with the objective to optimally regulate the state $\bar{x}(t)$ to $0$ with the minimum control effort $u(t)$. Following assumption is made on the system dynamics to facilitate the stability analysis.
\begin{assumption}
The function $g(x,\theta)$ is continuous and bounded $0<g(x,\theta)<\bar{g}$ with a known bound on a set $\mathcal{X} \subset \mathbb{R}^n$.
\end{assumption}

\section{Optimal Control Design using Actor-Critic Structure}
\subsection{Continuous RL-based Controller Design} \label{sec:ControlDesign}
An RL-based controller is designed to achieve the desired control objective given by the optimal value function $V^*(\bar{x}) \in \mathbb{R}^+$, defined by
\begin{equation}
V^*(\bar{x}(t)) = \textrm{min}_{u(\tau) \in \Theta(\mathcal{X})} \int_t^\infty r(s)ds  \label{eq:ValueFun}
\end{equation}
where $V^*$ is continuously differentiable, satisfies $V^*(0) = 0$, $\Theta(\mathcal{X})$ is a set of admissible policies, $r(\bar{x},u) \in \mathbb{R}$ is the local cost
\begin{equation}
    r(\bar{x},u) = Q(\bar{x}) + u^TRu  \label{eq:localCost}
\end{equation}
where $Q(\bar{x})$ is a positive definite function and $R = R^T >0$.
Given the dynamics (\ref{eq:ModSystem}) and the value function (\ref{eq:ValueFun}), the optimal control is given by 
\begin{equation}
    u^* = -\frac{1}{2} R^{-1}g^T(x,\theta)V^{*T}_{\bar{x}}
\end{equation}
where $V^*_{\bar{x}} = \frac{\partial V^*}{\partial \bar{x}}$.

\subsection{Hamiltonian and Bellman Error}
The Hamiltonian of the system is given by
\begin{equation}
    H(\bar{x},u,V_{\bar{x}}) = V_{\bar{x}}g(x,\theta)u + r(\bar{x},u)
\end{equation}
where $V_{\bar{x}} = \frac{\partial V}{\partial \bar{x}}$. The optimal Hamiltonian associated with the optimal cost and optimal control is given by
\begin{equation}
    H(\bar{x},u^*,V_{\bar{x}}^*) = V^*_{\bar{x}}g(x,\theta)u^* + r(\bar{x},u^*) = 0.
\end{equation}
Computing the value function $V(\bar{x})$ and the optimal controller requires the solution to the HJB which is a partial differential equation. It is, in general, hard to find the analytical solution of HJB. The value function is approximated using a NN called as a critic NN and the corresponding optimal control is approximated using an actor NN. Using the approximated cost and controller, the approximated Hamiltonian is computed as
\begin{equation}
    H(\bar{x},\hat{u},\hat{V}_{\bar{x}}) = \hat{V}_{\bar{x}}g(x,\hat{\theta})\hat{u} + r(\bar{x},\hat{u})
\end{equation}
Using the optimal and approximated Hamiltonian, a temporal difference or Bellman error $\delta \in \mathbb{R}$ is computed as follows
\begin{equation}
    \delta = H(\bar{x},\hat{u},\hat{V}_{\bar{x}}) - H(\bar{x},u^*,V_{\bar{x}}^*)  = \hat{V}_{\bar{x}}g(x,\hat{\theta})\hat{u} + r(\bar{x},\hat{u})  \label{eq:tempDiff}
\end{equation}
because the value of the optimal Hamiltonian is 0. Bellman error, $\delta$, in Hamiltonian is used to learn the critic and actor NN weights. For implementation of the optimal control, the value function and optimal control are approximated using NNs. Following assumptions are made on the NN function approximators.

\begin{assumption}
For a given NN, $N(x) = W^T\sigma(V^Tx) + \epsilon(x)$, where $x \in \mathcal{X}\subset \mathbb{R}^n$ is a compact set, $\epsilon(x)$ is a function reconstruction error, the ideal NN weights $W$ and $V$ are bounded by known positive constants, i.e., $\Vert W \Vert \leq \bar{W}$, $\Vert V \Vert \leq \bar{V}$ \cite{lewis2002neuro}. The NN activation function $\sigma$ and $\sigma'$ are bounded.
\end{assumption}
\begin{assumption}
Using the universal approximation property of NN, the function reconstruction error and its derivative are bounded, i.e., $\Vert \epsilon(x) \Vert \leq \bar{\epsilon}$ and $\Vert \epsilon'(x) \Vert \leq \bar{\epsilon}'$ \cite{cybenko1989approximation}.
\end{assumption}

\subsection{Approximate Optimal Control}
Consider a compact set $\mathcal{X} \subset \mathbb{R}^n$ and the state vector $\bar{x}(t) \in \mathcal{X}$. Using NN representation the optimal value function and the optimal control is written as
\begin{align}
    V^*(\bar{x}(t)) &= W_c^T \phi(\bar{x}) + \epsilon_c(\bar{x}) \nonumber \\
    u^*(\bar{x}) &= -\frac{1}{2}R^{-1}g^T(x,\theta)(\phi'(\bar{x})^T W_c + \epsilon_c'(\bar{x})^T) \label{eq:OptimalValueControl}
\end{align}
where $W_c \in \mathbb{R}^{n_c \times 1}$, $\phi(\bar{x}): \mathbb{R}^{n} \rightarrow \mathbb{R}^{n_c}$ are the basis functions, $\epsilon(\bar{x}) \in \mathbb{R}$ is the function error and $\epsilon'(\bar{x}) \in \mathbb{R}$ is its derivative with respect to $\bar{x}$.
Due to the function approximation error and unknown parameter in $g(x,\theta)$, the value function and optimal control cannot be implemented in practice. Thus, the approximated value function and the optimal control laws are designed as
\begin{align}
    \hat{V}(\bar{x}(t)) &= \hat{W}_c^T \phi(\bar{x}) \nonumber \\
    \hat{u}(\bar{x}) &= -\frac{1}{2}R^{-1}g^T(x,\hat{\theta})(\phi'(\bar{x})^T \hat{W}_a) \label{eq:ApproxValueControl}
\end{align}
where $\hat{W}_c \in \mathbb{R}^{n_c\times 1}$ and $\hat{W}_a \in \mathbb{R}^{n_a\times 1}$ are the estimated critic and actor weights. 

\subsection{Parameter Update Law}
To estimate the unknown parameter $\theta$ of $g(x,\theta)$, a CL-based parameter update law \cite{chowdhary2013concurrent} is designed as follows
\begin{align}
    \dot{\hat{\theta}} = &\gamma_\theta Y^{T}(\hat{u}^T \otimes \hat{W}_c^T \phi')^T + k_{cl}\gamma_\theta \sum_{j=1}^m \mathcal{Y}_i^T(\dot{\hat{x}}_j - \mathcal{Y}_i\hat{\theta}) \label{eq:thetaHatDot}
\end{align}
where $\mathcal{Y}_i = (\hat{u}_j^T \otimes I_{n\times n})Y_j$, $\gamma_\theta$, and $k_{cl}$ are constant gains. To implement the CL-based parameter update law in (\ref{eq:thetaHatDot}), a history stack is collected $\mathcal{H} = \{x_j,\hat{u}_j,\dot{\hat{x}}_j\}_{j=1}^{j=m}$ with $m$ number of data points. An estimate of the state derivative $\dot{\hat{x}}_j$ can be obtained using numerical techniques \cite{kamalapurkar2016model}. By collecting a history stack, information about the constant parameter $\theta$ can be obtained. CL-based parameter estimation law uses finite excitation condition of the system trajectories. Consider the parameter estimation error $\tilde{\theta}(t) = \theta - \hat{\theta}(t)$. Substituting $\dot{x}_j$ from (\ref{eq:SystemLIPForm}) in (\ref{eq:thetaHatDot}), the parameter estimation error dynamics can be written as 
\begin{equation}
    \dot{\tilde{\theta}} = -\gamma_\theta Y^{T}(\hat{u}^T \otimes \hat{W}_c^T \phi')^T - k_{cl}\gamma_\theta \sum_{j=1}^m \mathcal{Y}_i^T\mathcal{Y}_i\tilde{\theta} \label{eq:thetaTildeDot}
\end{equation}
%\begin{assumption}
%    The parameter estimate norm $\Vert \hat{\theta}(t) \Vert \neq %0$ for any time $t$.
%\end{assumption}
\begin{assumption} \label{ass:FiniteExcitation}
    For the history stack $\mathcal{H} = \{x_j,\hat{u}_j,\dot{\hat{x}}_j\}_{j=1}^{j=m}$ the following condition is satisfies
    $\lambda_{min}(\sum_{j=1}^m \mathcal{Y}_i^T \mathcal{Y}_i) = \sigma_1 > 0$,
    where $\sigma_1 \in \mathbb{R}^+$. The numerically computed derivatives of $x(t)$, $\dot{\hat{x}}_j$ computed at $j$th data point satisfies $\Vert \dot{\hat{x}}_j - \dot{x}_j \Vert \leq \epsilon$ for  a small positive number $\epsilon \in \mathbb{R}^+$. 
\end{assumption}
\begin{remark}
    Assumption \ref{ass:FiniteExcitation} is a finite excitation condition which can be verified in real-time \cite{chowdhary2013concurrent}.
\end{remark}
\begin{remark}
    The Initial excitation technique proposed in \cite{basu2019composite} or integral concurrent learning proposed in \cite{parikh2019integral} for parameter estimation in adaptive control can also be used for designing parameter update law.
\end{remark}
\begin{remark}
To avoid loss of control or to ensure $\dot{x} \neq 0$ for non-zero control $u$, a smooth projection operator $\mathrm{proj(\cdot)}$ is used on the parameter estimate update law to ensure that the estimates are projected to a region away from zero \cite{ioannou1996robust}.
\end{remark}
\begin{remark}
For $\theta \in \mathbb{R}$, the parameter update law reduces to $\dot{\hat{\theta}} = \gamma_\theta\hat{u}^TY^T\phi'^T\hat{W}_c +k_{cl}\gamma_\theta \sum_{j=1}^m u_j^TY_{rj}^T(\dot{x}_j - Y_{rj}u_j \hat{\theta})$
\end{remark}

\subsection{Bellman Error}
Let the actor-critic NN approximation errors be defined as $\tilde{W}_c(t) = W_c - \hat{W}_c(t)$ and $\tilde{W}_a(t) = W_a - \hat{W}_a(t)$. The actor and critic NN weights are updated using weight update laws that minimize the error between the approximated Hamiltonian and the optimal one, given by Bellman error. The Bellman error in a measurable form in terms of actor and critic NN weights is written as
\begin{align}
    \delta = \hat{W}_c^T \phi'(\bar{x}) g(x,\hat{\theta})\hat{u} + r(\bar{x},\hat{u})
\end{align}
For the analysis, another form of Bellman error based on (\ref{eq:tempDiff}) is derived as follows
\begin{align}
    \delta &= \hat{W}_c^T \phi'(\bar{x}) g(x,\hat{\theta})\hat{u} + \hat{u}^{T}R\hat{u} \nonumber \\
    &- W_c^{T} \phi'(\bar{x}) g(x,\theta)u^* - u^{*T}Ru^* - \epsilon_c' g(x,\theta) u^*
\end{align}
which by adding and subtracting $W_c^T \phi'(\bar{x}) g(x,\hat{\theta})\hat{u}$ can be written as
\begin{align}
    \delta &= -\tilde{W}_c^T w - W_c^T \phi'\tilde{g}\tilde{u} -\epsilon_c' gu^* + \frac{1}{4}\tilde{W}_a^Tg_\phi\tilde{W}_a \nonumber \\
    & -\frac{1}{2}\tilde{W}_a^T g_\phi W_c +\frac{1}{4}\hat{W}^T_a \tilde{g}_\phi \hat{W}_a -\frac{1}{2}\hat{W}_a^T \tilde{g}_{a\phi} \hat{W}_a \nonumber \\
    &-\frac{1}{4}\epsilon_c' g_r \epsilon_c'^T  -\frac{1}{2}\epsilon_c'^T g_r \phi'^T W_c \label{eq:delta}
\end{align}
where $g_\phi=\phi'gR^{-1}g^T\phi'^T$, $g_{r} = g(x,\theta)R^{-1}g(x,\theta)^{T}$, $\tilde{g}_{a\phi} = \phi'gR^{-1}\tilde{g}^T\phi'^T$, $\tilde{g}_\phi = \phi' \tilde{g} R^{-1} \tilde{g}^T \phi'^T$, $\hat{a}^TR\hat{a}-a^{*T}Ra^* = \tilde{a}^T R \tilde{a} - 2\tilde{a}^T R a^*$ is used for $\hat{u}^{T}R\hat{u}-u^{*T}Ru^*$ and $w \in \mathbb{R}^{n_c}$ is defined as
\begin{equation}
    w = \phi'(\bar{x})g(x,\hat{\theta})\hat{u}.   \label{eq:w}
\end{equation} 
\subsection{Critic NN Weight Update Law}
The least-squares update law for the critic NN can be derived by minimizing the integral Bellman error $E_{cr} = \int_0^t \frac{1}{2}\delta^2(\tau) d\tau$.
Taking the time derivative of $E_{cr}$ with respect to $\hat{W}_c$
\begin{equation}
    \frac{\partial E_{cr}}{\partial \hat{W}_c} = 2 \int_0^t \delta  \frac{\partial \delta}{\partial \hat{W}_c} d\tau = 0
\end{equation}
where $\frac{\partial \delta}{\partial \hat{W}_c}= w^T$.
%\begin{equation}
%    \frac{\partial E_{cr}}{\partial \hat{W}_c} = 2 \int_0^t \hat{W}_c^T ww^T + %r(x,\hat{u})w^T d\tau = 0
%\end{equation}
The least squares solution can be derived as
\begin{equation}
    \dot{\hat{W}}_c = \mathrm{proj}( - \gamma_c \Gamma \Omega \delta ) \label{eq:wCHatUpdate}
\end{equation}
where $\Omega = \frac{w}{1 + c_1 w^T\Gamma w} \in \mathbb{R}$, $c_1 \in \mathbb{R}$ and $\gamma_c \in \mathbb{R}$ are constant gains, $\Gamma(t) = (\int_0^t w(\tau)w(\tau)^T d\tau)^{-1}$ $\in \mathbb{R}^{n_c \times n_c} $ is a symmetric estimation gain matrix, which is computed using
\begin{equation}
\dot{\Gamma} = -\gamma \Gamma \frac{ww^T}{1 + c_1w^T\Gamma w}\Gamma, \qquad \Gamma(0) = c_2I \label{eq:GammaUpdate}
\end{equation}
where $\gamma \in \mathbb{R}_+$ and $c_2 \in \mathbb{R}^+$.
\begin{assumption} \label{ass:assumptionPE}
    The normalized critic regressor $\xi = \frac{w}{\sqrt{1 + c_1 w^T\Gamma w}}$ is bounded and is persistently exciting (PE), i.e.,    \begin{equation}
      \mu_a I \leq \int_{t_0}^{t_0+T} \xi(\tau) \xi^T(\tau) d\tau \leq \mu_b I, \quad \forall t_0>0
    \end{equation}
    where $\mu_a$, $\mu_b$, $T$ are positive constants. \cite{bhasin2013novel}
\end{assumption}
Consider the error in the critic weights $\tilde{W}_c = W_c - \hat{W}_c$. Taking derivative of $\tilde{W}_c$ results in
\begin{align}
\dot{\tilde{W}}_{c}&=-\gamma_{c} \Gamma \xi \xi^T \tilde{W_{c}}+\gamma_{c}\Omega \left(-W_{c}^{T}\phi'\tilde{g}\tilde{u}+\frac{1}{4}\tilde{W}_{a}^{T}g_{\phi}\tilde{W_{a}} \right. \nonumber\\
&-\frac{1}{2}\tilde{W_{a}^{T}}g_{\phi}W_{c} -\epsilon_{c}^{'}gu^{*}-\frac{1}{4}\epsilon_{c}^{'}g_{r}\epsilon_{c}^{'T}-\frac{1}{2}\epsilon_{c}^{'T}g_{r}\phi'^{T}W_{c} \nonumber \\
&\left. +\frac{1}{4}\hat{W}_a^T \tilde{g}_\phi \hat{W}_a - \frac{1}{2} \hat{W}_a^T \tilde{g}_{a\phi} \hat{W}_a \right) \label{eq:wctildeDot}
\end{align}
Under Assumption \ref{ass:assumptionPE}, a nominal system formed using first term of (\ref{eq:wctildeDot}) is globally exponentially stable \cite{sastry1990adaptive,bhasin2013novel}, which according to converse Lyapunov Theorem induces a Lyapunov function $V_{wc}(t,\tilde{W}_c)$ with following properties 
\begin{align}
\gamma_1 \Vert \tilde{W}_c \Vert^2 \leq V_{wc}(t,\tilde{W}_c) &\leq \gamma_2 \Vert \tilde{W}_c \Vert^2 \nonumber\\
\frac{\partial {V}_{wc}}{\partial t} + \frac{\partial V_{wc}}{\partial \tilde{W}_{wc}} (-\gamma_{c} \Gamma \xi \xi^T \tilde{W_{c}}) &\leq -\eta_1 \Vert \tilde{W}_c\Vert^2 \nonumber \\
\Vert \frac{\partial V_{wc}}{\partial \tilde{W}_c}\Vert &\leq \bar{\gamma} \Vert \tilde{W}_c \Vert \label{eq:ConverseLyapunov}
\end{align}
where  $\gamma_1, \gamma_2, \eta_1, \bar{\gamma} \in \mathbb{R}^+$.

\subsection{Actor NN Weight Update Law}
The least-squares gradient-based update law for the actor NN is derived using the squared Bellman error $E_a = \delta^2$. Computing the gradient of $E_a$ and setting it to zero, results in the following update law for $\hat{W}_a$ 
\begin{equation}
\dot{\hat{W}}_a = \mathrm{proj} \left(-\frac{\gamma_a \hat{g}_\phi (\hat{W}_a - \hat{W}_c)\delta}{\sqrt{1+w^Tw}} - \gamma_{a2}(\hat{W}_a - \hat{W}_c) \right) \label{eq:wAHatUpdate}
\end{equation}
where $\hat{g}_\phi = \phi'g(x,\hat{\theta})R^{-1}g^T(x,\hat{\theta})\phi'^T$, $\gamma_a$ and $\gamma_{a2}$ are constant gains. For the stability analysis presented in next section, following bounds are defined
\begin{gather}
\Vert\frac{1}{4}\epsilon'_{c}g_{r}\epsilon_{c}^{'T}-\frac{1}{4}W_{c}^{T}g_{\phi}W_{c}+\frac{1}{2}\epsilon'_{c}g_{r}W_{c} \nonumber \\ -\frac{1}{2}W_c^T \tilde{g}_\phi W_c - \frac{1}{2}\epsilon_c'g_r\phi'^T W_a \Vert \leq \kappa_{1}  \label{eq:bound1} \\
\Vert\frac{1}{2}W_{c}^{T}g_{\phi}\Vert\leq\kappa_{2}, \quad\Vert\frac{1}{2}\epsilon_{c}'g_{r}\phi'^{T}\Vert\leq\kappa_{3} \label{eq:bound2} \\
\Vert\frac{1}{4}\tilde{W}_{a}^{T}g_{\phi}\tilde{W}_{a}\Vert\leq\kappa_{4}, \quad \overline{w}=\left\Vert g_{r}\phi'^{T}(\hat{W}_{a}-\hat{W}_{c})\right\Vert \label{eq:bound3} \\
    \Vert-W_{c}^{T}\phi'\tilde{g}\tilde{u}-\epsilon_{c}'gu^{*}-\frac{1}{4}\epsilon'_{c}g_{r}\epsilon_{c}^{'T}-\frac{1}{2}\epsilon_{c}^{'T}g_{r}\phi^{'T}W_{c} \nonumber \\
    +\frac{1}{4}\hat{W}_a^T \tilde{g}_\phi \hat{W}_a - \frac{1}{2} \hat{W}_a^T \tilde{g}_{a\phi} \hat{W}_a\Vert\leq\kappa_{5} \label{eq:bound4}
\end{gather}
where $\kappa_1$, $\kappa_2$, $\kappa_3$, $\kappa_4$, $\kappa_5$ and $\bar{\omega}$ are positive constants.
\section{Stability Analysis}
\begin{theorem}
Given that the Assumptions 1-5 hold and the following sufficient condition is satisfied 
\begin{equation}
    \gamma_{a2} - \gamma_c\kappa_{4}\overline{w} > 0,
\end{equation}
the actor-critic controller (\ref{eq:ApproxValueControl}) along with the model parameter update law in (\ref{eq:thetaHatDot}) and critic and actor weight update laws in (\ref{eq:wCHatUpdate})-(\ref{eq:GammaUpdate}), (\ref{eq:wAHatUpdate}) guarantee that the signals $\bar{x}(t)$, $\tilde{\theta}(t)$, $\tilde{W}_a(t)$ and $\tilde{W}_c(t)$ are uniformly ultimately bounded.
\end{theorem}
\begin{proof}
Consider a positive definite continuously differentiable Lyapunov function $V:\mathcal{X} \times \mathbb{R}^{n_c} \times \mathbb{R}^{n_a} \times \mathbb{R} \times [0,\infty) \rightarrow \mathbb{R}^+$
\begin{equation}
    V_z = V^*(t) + V_{wc}(t,\tilde{W}_c) \nonumber + \frac{1}{2}\tilde{W}_a^T \tilde{W}_a + \frac{1}{2\gamma_\theta}\tilde{\theta}^T\tilde{\theta} \label{eq:LyapunovCandidate}
\end{equation}
where $V^*(t)$ is the optimal value function, $V_{wc}$ is a Lyapunov function defined in (\ref{eq:ConverseLyapunov}), $\gamma_\theta \in \mathbb{R}^+$. Since the optimal value function $V^*(t)$ is continuously differentiable and positive definite, there exists class-$\mathcal{K}$ functions $\alpha_1(\cdot)$ and $\alpha_2(\cdot)$ such that
%\begin{equation}
$\alpha_1(\Vert \bar{x} \Vert ) \leq V^*(t) \leq \alpha_2(\Vert \bar{x} \Vert), \; \forall \bar{x} \in \mathcal{B}_a \subset \mathcal{X}$. %\label{eq:VStarBound}
%\end{equation}
Let us define $z(t) = [\bar{x}(t)^T, \tilde{W}_c(t)^T,\tilde{W}_a(t)^T, \tilde{\theta}]^T\in \mathbb{R}^{n+n_c+n_a+p}$. Based on the bounds on $V^*(t)$, the following bounds can be derived
\begin{equation}
    \alpha_3(\Vert z \Vert ) \leq V_z(\bar{x},\tilde{W}_c, \tilde{W}_a,\tilde{\theta},t) \leq \alpha_4(\Vert z \Vert) \quad \forall \bar{x} \in \mathcal{B}_z \label{eq:VzBound}
\end{equation}
where $\alpha_3(\cdot)$ and $\alpha_4(\cdot)$ are class $\mathcal{K}$-functions.
Taking the time derivative of the Lyapunov function and
%\begin{align}
%    \dot{V}_z &= V^*_{\bar{x}} g(x,\theta)\hat{u} + \dot{V}_{wc} + \tilde{W}_a^T\dot{\tilde{W}}_a + \frac{1}{\gamma_\theta}\tilde{\theta}^T\dot{\tilde{\theta}}
%\end{align}
using $\dot{\tilde{W}}_a = -\dot{\hat{W}}_a$ results in
\begin{align}
    \dot{V}_z = V^*_{\bar{x}} g(x,\theta)\hat{u} - \tilde{W}_a^T\dot{\hat{W}}_a + \dot{V}_{wc} + \frac{1}{\gamma_\theta}\tilde{\theta}^T\dot{\tilde{\theta}}
\end{align}
Adding and subtracting $V^*_{\bar{x}} g(x,\theta)u^*$,
%\begin{align}
%    \dot{V} &= V^*_{\bar{x}} g(x,\theta)u^* - V^*_{\bar{x}} g(x,\theta) (u^*-\hat{u})  \nonumber \\
%    & + \dot{V}_{wc} - \tilde{W}_a^T\dot{\hat{W}}_a %-\frac{\partial V^*}{\partial x}Y_r(x)\hat{u}\tilde{\theta} 
%    - \frac{1}{\gamma_\theta}\tilde{\theta}^T\dot{\hat{\theta}}
%\end{align}
utilizing $V^*_{\bar{x}}g(x,\theta) = -2u^{*T}R$ and $V^*_{\bar{x}}g(x,\theta)u^* = -Q(\bar{x}) - u^{*T}Ru^*$, and
%\begin{align}
%    \dot{V} &= -Q(\bar{x}) - u^{*T}Ru^* + 2u^{*T}R(u^*-\hat{u})  \nonumber \\
%    &  + \dot{V}_{wc}  - \tilde{W}_a^T\dot{\hat{W}}_a 
    %-\frac{\partial V^*}{\partial x}Y_r(x)\hat{u}\tilde{\theta} \nonumber \\
%     - \frac{1}{\gamma_\theta}\tilde{\theta}^T\dot{\hat{\theta}}
%\end{align}
substituting $u^*$ from (\ref{eq:OptimalValueControl}), $\hat{u}$ from (\ref{eq:ApproxValueControl}),
%\begin{align}
%    \dot{V}_z &= -Q(\bar{x}) + \frac{1}{4}V^*_{\bar{x}}g(x,\theta)R^{-1}g(x,\theta)^TV_{\bar{x}}^{*T} + \dot{V}_{wc} \nonumber \\
%    & - \frac{1}{2}V^*_{\bar{x}}g(x,\theta)R^{-1}g^T(x,\hat{\theta})\phi^{'T}\hat{W}_a  - \tilde{W}_a^T\dot{\hat{W}}_a 
%    + \frac{1}{\gamma_\theta}\tilde{\theta}^T\dot{\tilde{\theta}}
%\end{align}
the NN form of the $V^*$ from (\ref{eq:OptimalValueControl}) and (\ref{eq:thetaTildeDot}) results in
\begin{align}
\dot{V}_z&=-Q(\bar{x})+\frac{1}{4}\epsilon'_{c}g_{r}\epsilon^{'T}_{c}+\frac{1}{4}W_{c}^{T}g_{\phi}W_{c} +\frac{1}{2}\epsilon'_{c}g_{r}W_{c}\nonumber \\
&-\frac{1}{2}W_{c}^{T}\phi'gR^{-1}\hat{g}^T\phi'^T\hat{W}_a-\frac{1}{2}\epsilon_{c}'gR^{-1}\hat{g}^T\phi'^{T}\hat{W}_{a} + \dot{V}_{wc} \nonumber \\
&-\tilde{W}_{a}^{T}\dot{\hat{W}}_{a}-\tilde{\theta}^{T}Y^{T}(\hat{u}^T \otimes \hat{W}_c^T \phi')^T - k_{cl} \tilde{\theta} \sum_{j=1}^m \mathcal{Y}_i^T\mathcal{Y}_i\tilde{\theta} %\label{eq:VDot}
\end{align}
Adding and subtracting $W_c^T\phi'\hat{g}\hat{u}$ and $\hat{W}_c^T\phi'\tilde{g}\hat{u}$, the following expression is obtained
%\begin{align}
%\dot{V}&=-Q(\bar{x})+\frac{1}{4}\epsilon'_{c}g_{r}\epsilon_{c}^{'T}
%+\frac{1}{4}W_{c}^{T}g_{\phi}W_{c}+\frac{1}{2}\epsilon'_{c}g_{r}W_{c} \nonumber \\
%&-\frac{1}{2}\epsilon_{c}'g_{r}\phi'^{T}\hat{W}_{a} 
%+W_c^T\phi'\tilde{g}\hat{u} -\frac{1}{2}W_c^T \hat{g}_\phi W_c + \dot{V}_{wc} \nonumber \\
%&-\tilde{W}_{a}^{T}\dot{\hat{W}}_{a}-\tilde{\theta}^{T}Y_{r}^{T}(\hat{u}^T \otimes \hat{W}_c^T \phi')^T - k_{cl} \tilde{\theta} \sum_{j=1}^m \mathcal{Y}_i^T\mathcal{Y}_i\tilde{\theta}  %\label{eq:VDot}
%\end{align}
%which by adding and subtracting $\hat{W}_c^T\phi'\tilde{g}\hat{u}$ can further be written as
\begin{align}
\dot{V}_z&=-Q(\bar{x})+\frac{1}{4}\epsilon'_{c}g_{r}\epsilon_{c}^{'T}
+\frac{1}{4}W_{c}^{T}g_{\phi}W_{c}+\frac{1}{2}\epsilon'_{c}g_{r}W_{c} \nonumber \\
&-\frac{1}{2}\epsilon_{c}'g_{r}\phi'^{T}\hat{W}_{a} 
+\tilde{W}_{c}^{T}\phi'\tilde{g}\hat{u} + \hat{W}_{c}^{T}\phi'\tilde{g}\hat{u} \nonumber \\
&-\frac{1}{2}W_c^T \hat{g}_\phi W_c + \dot{V}_{wc} -\tilde{W}_{a}^{T}\dot{\hat{W}}_{a}
-\tilde{\theta}^{T}Y_{r}^{T}(\hat{u}^T \otimes \hat{W}_c^T \phi')^T  \nonumber \\
&- k_{cl} \tilde{\theta} \sum_{j=1}^m \mathcal{Y}_i^T\mathcal{Y}_i\tilde{\theta} %\label{eq:VDot}
\end{align}
Using the fact that $\hat{W}_c^T\phi'\tilde{g}\hat{u} = (\hat{u}^T \otimes \hat{W}_c^T \phi')Y\tilde{\theta}$
\begin{align}
\dot{V}_z&=-Q(\bar{x})+\frac{1}{4}\epsilon'_{c}g_{r}\epsilon_{c}^{'T}
+\frac{1}{4}W_{c}^{T}g_{\phi}W_{c}+\frac{1}{2}\epsilon'_{c}g_{r}W_{c} \nonumber \\
&-\frac{1}{2}\epsilon_{c}'g_{r}\phi'^{T}\hat{W}_{a} 
+\tilde{W}_{c}^{T}\phi'\tilde{g}\hat{u} + (\hat{u}^T \otimes \hat{W}_c^T \phi')Y\tilde{\theta} \nonumber \\
&-\frac{1}{2}W_c^T \hat{g}_\phi W_c + \dot{V}_{wc} -\tilde{W}_{a}^{T}\dot{\hat{W}}_{a}-\tilde{\theta}^{T}Y^{T}(\hat{u}^T \otimes \hat{W}_c^T \phi')^T  \nonumber \\
&- k_{cl} \tilde{\theta} \sum_{j=1}^m \mathcal{Y}_i^T\mathcal{Y}_i\tilde{\theta}  \label{eq:VDot}
\end{align}
To simplify the $\dot{V}_z$ expression in (\ref{eq:VDot}), consider $-\tilde{W}_a^T\dot{\hat{W}}_a$ after substituting actor weight update law from (\ref{eq:wAHatUpdate})
%\begin{align}
%-\tilde{W}_{a}^{T}\dot{\hat{W}}_{a} &=-C_{1}\tilde{W}_{a}^{T}\hat{g}_{\phi} \hat{W}_{c}\delta + C_{1}\tilde{W}_{a}^{T}\hat{g}_{\phi}\hat{W_{a}}\delta \nonumber \\
%&+\gamma_{a2}\tilde{W}_{a}^{T}(\hat{W}_{a}-\hat{W}_{c}) \label{eq:whatDotTerm}
%\end{align}
%The term in (\ref{eq:whatDotTerm}) can be written as
\begin{align}
-\tilde{W}_{a}^{T}\dot{\hat{W}}_{a} & =C_{1}\tilde{W}_{a}^{T}\left(\hat{g}_{\phi}(\hat{W}_{a}-\hat{W}_{c})\right)\delta \nonumber \\
&-\gamma_{a2}\tilde{W}_{a}^{T}\tilde{W}_{a}+\gamma_{a2}\tilde{W}_{a}^{T}\tilde{W}_{c} \label{eq:wahatDotTerm}
\end{align}
where $C_1 = \frac{\gamma_a}{\sqrt{1+w^Tw}}$. 
%Consider $\dot{\tilde{W}}_c$ term after substituting $\delta$ from (\ref{eq:delta})
%\begin{align}\dot{\tilde{W}}_{c}&=-\gamma_{c}\Omega w^{T}\tilde{W_{c}}+\gamma_{c}\Omega \nonumber \\
%&\left(-W_{c}^{T}\phi'\tilde{g}\tilde{u}+\frac{1}{4}\tilde{W}_{a}^{T}g_{\phi}\tilde{W_{a}}-\frac{1}{2}\tilde{W_{a}^{T}}g_{\phi}W_{c}\right. \nonumber \\
%&-\epsilon_{c}^{'}gu^{*}-\frac{1}{4}\epsilon_{c}^{'}g_{r}\epsilon_{c}^{'T}-\frac{1}{2}\epsilon_{c}^{'T}g_{r}\phi'^{T}W_{c} \nonumber \\
%&\left. \textcolor{blue}{ +\frac{1}{4}\hat{W}_a^T \tilde{g}_\phi \hat{W}_a - \frac{1}{2} \hat{W}_a^T \tilde{g}_{a\phi} \hat{W}_a} \right) \label{eq:wctildeDot}
%\end{align}
Substituting (\ref{eq:wahatDotTerm}) and the bounds on Lyapunov function $V_{wc}$ from (\ref{eq:ConverseLyapunov}) into (\ref{eq:VDot}), adding and subtracting terms $\frac{1}{2}W_c^Tg_{\phi}W_c$ and $\frac{1}{2}\epsilon_c'g_r\phi'^T W_a$, $\dot{V}_z$ can be written as
\begin{align}
\dot{V}_z&=-Q(\bar{x})-\eta_{1}\Vert\tilde{W}_{c}\Vert^{2}-\gamma_{a2}\Vert\tilde{W}_{a}\Vert^{2} +\frac{1}{4}\epsilon'_{c}g_{r}\epsilon_{c}^{'T} \nonumber \\ 
&- \frac{1}{2}\epsilon_c'g_r\phi'^T W_a 
-\frac{1}{4}W_{c}^{T}g_{\phi}W_{c}+\frac{1}{2}\epsilon'_{c}g_{r}W_{c}\nonumber \\
&+\frac{1}{2}\epsilon_{c}'g_{r}\phi'^{T}\tilde{W}_{a}
-\frac{1}{2}W_c^T \tilde{g}_\phi W_c +\bar{\gamma}\Vert \tilde{W}_{c}\Vert\Bigg(-W_{c}^{T}\phi'\tilde{g}\tilde{u}\nonumber \\
&+\frac{1}{4}\tilde{W}_{a}^{T}g_{\phi}\tilde{W}_{a} -\frac{1}{2}\tilde{W}_{a}^{T}g_{\phi}W_{c} -\epsilon_{c}'gu^{*}-\frac{1}{4}\epsilon'_{c}g_{r}\epsilon_{c}^{'T}
\nonumber \\
&-\frac{1}{2}\epsilon_{c}^{'T}g_{r}\phi^{'T}W_{c} +\frac{1}{4}\hat{W}_a^T \tilde{g}_\phi \hat{W}_a - \frac{1}{2} \hat{W}_a^T \tilde{g}_{a\phi} \hat{W}_a\Bigg) \nonumber \\
&- k_{cl} \tilde{\theta} \sum_{j=1}^m \mathcal{Y}_i^T\mathcal{Y}_i\tilde{\theta} +\gamma_{a2}\tilde{W}_{a}^{T}(W_{c}-\hat{W}_{c}) \nonumber \\
&+C_{1}\tilde{W}_{a}^{T}\left(\hat{g}_{\phi}(\hat{W}_{a}-\hat{W}_{c})\right)\delta 
\end{align} 
Utilizing bounds in (\ref{eq:bound1})-(\ref{eq:bound4}), using $\vert C_1 \vert \leq \gamma_c$, and completing the squares, $\dot{V}_z$ can be upper bounded as  
%\begin{align}
%\dot{V}&\leq-Q(\bar{x})-\eta_{1}\Vert\tilde{W}_{c}\Vert^{2}-2C_{2}\Vert\tilde{W}_{a}\Vert^{2}+\kappa_{1} \nonumber\\
%&+\kappa_{3}\Vert\tilde{W}_{a}\Vert+\bar{\gamma}\kappa_{5}\Vert\tilde{W}_{c}\Vert+\frac{1}{4}\bar{\gamma}\kappa_{4}\Vert\tilde{W}_{c}\Vert \nonumber\\
%&+\frac{1}{2}\bar{\gamma}\kappa_{2}\Vert\tilde{W}_{c}\Vert\tilde{W}_{a}\Vert -C_{1}\tilde{W}_{a}^{T}\left(g_{\phi}(\hat{W}_{a}-\hat{W}_{c})\right)\delta \nonumber\\
%&+C_{2}\Vert\tilde{W}_{a}\Vert\Vert\tilde{W}_{c}\Vert - k_{cl} \tilde{\theta} \sum_{j=1}^m \mathcal{Y}_i^T\mathcal{Y}_i\tilde{\theta}
%\end{align} 
%\begin{align}
%    \dot{V}&\leq-Q(\bar{x})-\eta_{1}\Vert\tilde{W}_{c}\Vert^{2}-2C_2\Vert\tilde{W}_{a}\Vert^{2} -\sigma_1\Vert \tilde{\theta}\Vert ^2+\kappa_{1} \nonumber\\   &+\kappa_{3}\Vert\tilde{W}_{a}\Vert+\bar{\gamma}\kappa_{5}\Vert\tilde{W}_{c}\Vert+\frac{1}{4}\bar{\gamma}\kappa_{4}\Vert\tilde{W}_{c}\Vert \nonumber\\
%    &+\frac{1}{2}\bar{\gamma}\kappa_{2}\Vert\tilde{W}_{c}\Vert\tilde{W}_{a}\Vert+C_{1}\kappa_{5}\overline{w}_{j}\Vert \tilde{W}_{a}\Vert +C_{1}\kappa_{4}\overline{w}_{j}\Vert \tilde{W}_{a}\Vert ^{2} \nonumber \\
%    &+C_{1}\kappa_{2}\overline{w}_{j}\Vert \tilde{W}_{a}\Vert +C_{2}\Vert\tilde{W}_{a}\Vert\Vert\tilde{W}_{c}\Vert 
%\end{align}
\begin{align}
\dot{V}_z & \leq -Q(\bar{x})-(1-\theta_{1})\eta_{1}\Vert\tilde{W}_{c}\Vert^{2} \nonumber\\
&-(1-\theta_{1})\left(2\gamma_{a2}-\gamma_c\kappa_{4}\overline{w}\right)\Vert\tilde{W}_{a}\Vert^{2} - \sigma_1\Vert \tilde{\theta}\Vert ^2 \nonumber\\
&+\kappa_{1}+(\kappa_{2}+\kappa_{3}+C_{1}(\kappa_{5}+\kappa_{2})\overline{w}_{j})\Vert\tilde{W}_{a}\Vert \nonumber\\
&+\left(\bar{\gamma}\kappa_{5} +\frac{1}{4}\bar{\gamma}\kappa_{4}\right)\Vert\tilde{W}_{c}\Vert 
\end{align}
where $\eta_2 = \gamma_{a2} - \gamma_c\kappa_{4}\overline{w}$, $\sigma_1 = k_{cl} \lambda_\mathrm{min}(\sum_{j=1}^m \mathcal{Y}_i^T\mathcal{Y}_i)$, $1-\theta_1>0$, the final bound on $\dot{V}_z$ can derived as
\begin{align}
    \dot{V}_z &\leq-Q(\bar{x})-(1-\theta_{1}-\theta_{2})\eta_{1}\Vert\tilde{W}_{c}\Vert^{2} \nonumber \\
    &-(1-\theta_{1}-\theta_{2})\eta_{2}\Vert\tilde{W}_{a}\Vert^{2} - \sigma_1\Vert \tilde{\theta}\Vert ^2+\kappa_{1} \nonumber\\    &+\frac{(\bar{\gamma}\kappa_{5}+\frac{1}{4}\bar{\gamma}\kappa_{4})^{2}}{4(1-\theta_{1}-\theta_{2})\eta_{1}} +\frac{(\kappa_{2}+\kappa_{3}+C_{1}(\kappa_{5}+\kappa_{2})\overline{w})^{2}}{4(1-\theta_{1}-\theta_{2})\eta_{2}} \label{eq:VDotBound}
\end{align}
where $1-\theta_1 - \theta_2>0$. Since $Q(\bar{x})$ is positive definite, Lemma 4.3 of \cite{Khalil2002} can be utilized to derive
\begin{align}
    \alpha_5 (\Vert z \Vert) & \leq  Q + (1-\theta_{1}-\theta_{2})\eta_{1}\Vert\tilde{W}_{c}\Vert^{2} + \sigma_1\Vert \tilde{\theta}\Vert ^2
    \nonumber \\
    &+(1-\theta_{1}-\theta_{2})\eta_{2}\Vert\tilde{W}_{a}\Vert^{2} \leq \alpha_6 (\Vert z \Vert) \label{eq:classKFun}
\end{align}
where $\alpha_5(\cdot)$ and $\alpha_6(\cdot)$ are class-$\mathcal{K}$ functions.
Using (\ref{eq:classKFun}), the expression (\ref{eq:VDotBound}) can be upper bounded as
\begin{align}
    \dot{V}_z &\leq -\alpha_5(\Vert z \Vert ) + \kappa_{1} +\frac{(\bar{\gamma}\kappa_{5}+\frac{1}{4}\bar{\gamma}\kappa_{4})^{2}}{4(1-\theta_{1}-\theta_{2})\eta_{1}} \nonumber \\
    &+\frac{(\kappa_{2}+\kappa_{3}+C_{1}(\kappa_{5}+\kappa_{2})\overline{w})^{2}}{4(1-\theta_{1}-\theta_{2})\eta_{2}} \label{eq:VDotBound}
\end{align}
which proves that $\dot{V}_z$ is always negative whenever $z(t)$ is outside the compact set 
\begin{align}
    \bar{\Omega} &= \{z : \Vert z \Vert \leq \alpha_5^{-1} (\kappa_{1} +\frac{(\bar{\gamma}\kappa_{5}+\frac{1}{4}\bar{\gamma}\kappa_{4})^{2}}{4(1-\theta_{1}-\theta_{2})\eta_{1}} 
    \nonumber \\
    & +\frac{(\kappa_{2}+\kappa_{3}+C_{1}(\kappa_{5}+\kappa_{2})\overline{w})^{2}}{4(1-\theta_{1}-\theta_{2})\eta_{2}} )\}.
\end{align}
Invoking Theorem 4.18 of \cite{Khalil2002} $\Vert z \Vert$ is uniformly ultimately bounded (UUB). 
\end{proof}
\begin{remark}
    The ultimate bound on $\Vert z \Vert$ can be reduced by appropriately choosing the gains $\gamma_{a2}$, $\gamma_c$, and increasing the number of neurons in the NN which reduces the function approximation error of the actor and critic NN. 
\end{remark}
\begin{remark}
    When the Assumption \ref{ass:FiniteExcitation} is not satisfied then the term $\sigma_1\Vert \tilde{\theta} \Vert^2$ will not be there in $\dot{V}_z$ expression in (\ref{eq:VDotBound}). This is the time period when the history stack data is collected and $\sum_{j=1}^m \mathcal{Y}_i^T \mathcal{Y}_i$ is not full rank. To yield a UUB stability result a sigma modification term is added to the model parameter update law (\ref{eq:thetaHatDot}) \cite{ioannou1996robust}, which yields the parameter update law 
    \begin{equation}
        \dot{\hat{\theta}} = \gamma_\theta Y^{T}(\hat{u}^T \otimes \hat{W}_c^T \phi')^T - \sigma_2 \hat{\theta}, \;\; \mathrm{det}\{\sum_{j=1}^m \mathcal{Y}_i^T \mathcal{Y}_i\} = 0  \label{eq:ParameterUpdateSigmaMod}
    \end{equation}
    Since the finite excitation condition of Assumption \ref{ass:FiniteExcitation} can be verified in real time, the time of switching between (\ref{eq:ParameterUpdateSigmaMod}) and (\ref{eq:thetaHatDot}) can be computed.
\end{remark}
\begin{remark}
    Due to switching of the model parameter update law from (\ref{eq:ParameterUpdateSigmaMod}) to (\ref{eq:thetaHatDot}) based on the finite excitation condition, the error system of $z(t)$ is a switched system. The Lyapunov function in (\ref{eq:LyapunovCandidate}) is a common Lyapunov function for the error system.
\end{remark}
\section{Simulation Studies}
Simulations are carried out to test the performance of the RL-based controller on IBVS and WMR control.
\subsection{Optimal IBVS Controller}
\begin{figure*}
	   \begin{subfigure}{}
		\includegraphics[width=0.3\textwidth]{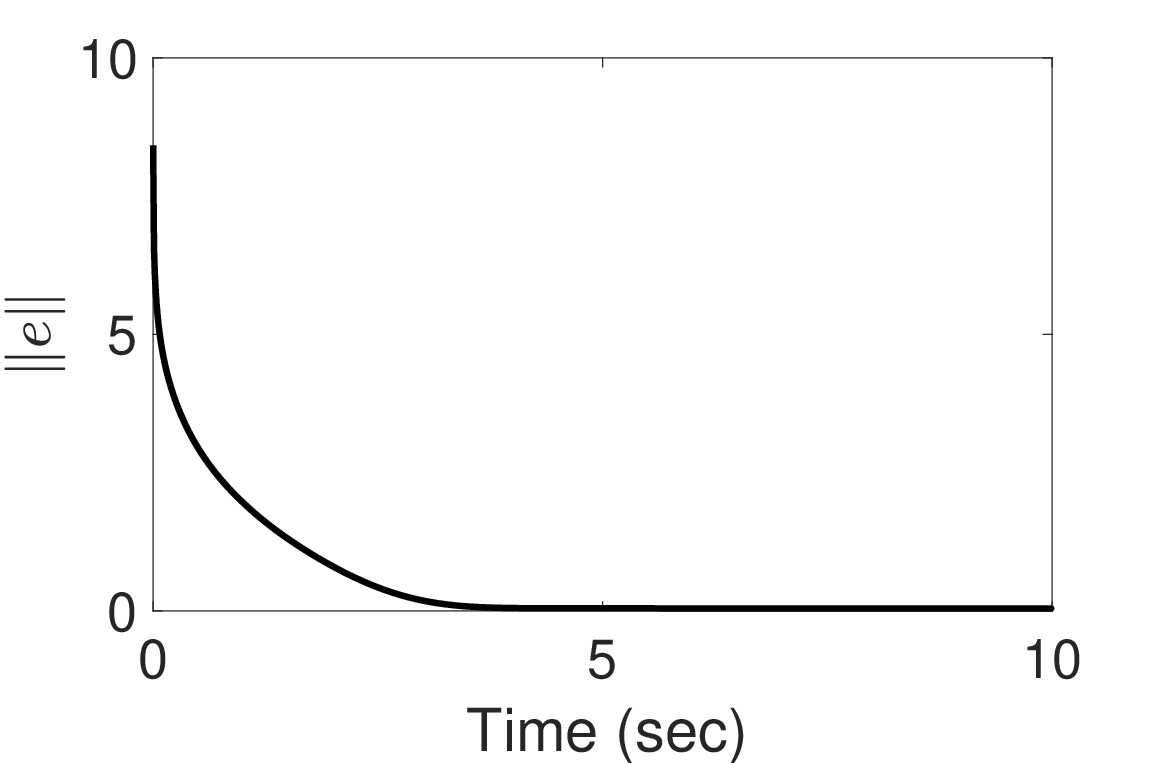}
		\label{fig:IBVSErrors}
	   \end{subfigure}
	   \begin{subfigure}{}
		\includegraphics[width=0.3\textwidth]{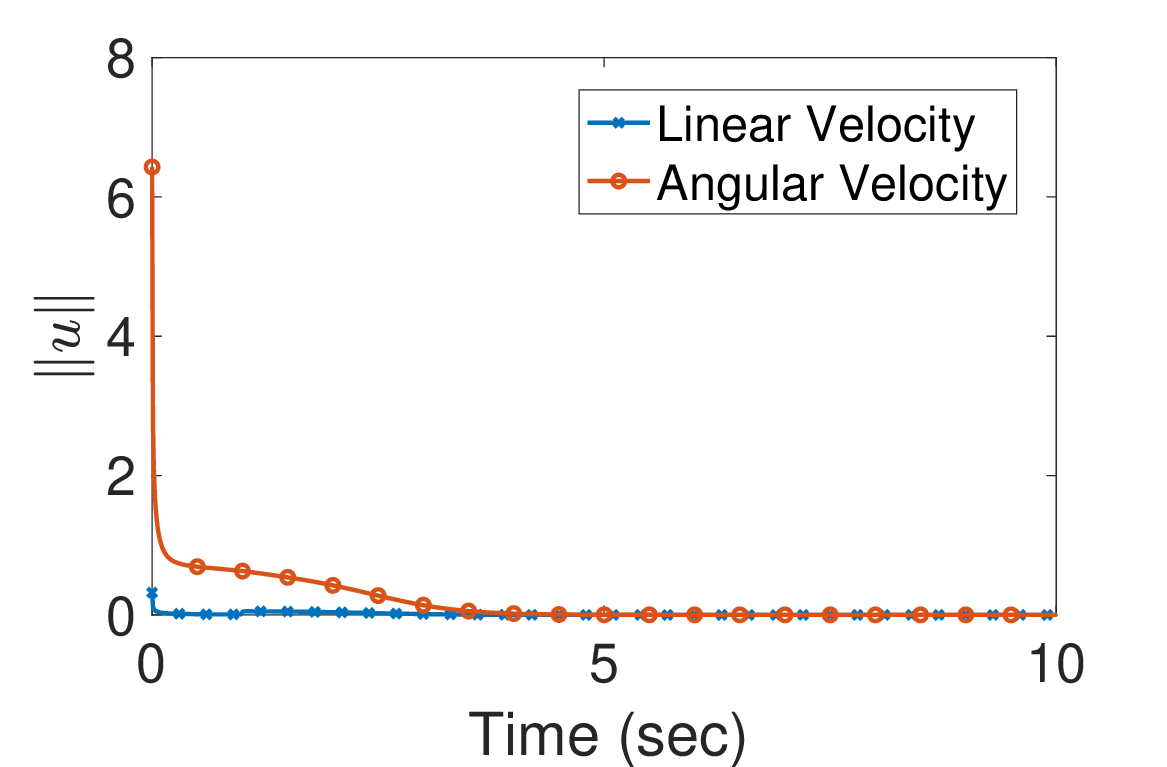}
		\label{fig:IBVSControl}
	    \end{subfigure}
	   \begin{subfigure}{}
		  \includegraphics[width=0.3\textwidth]{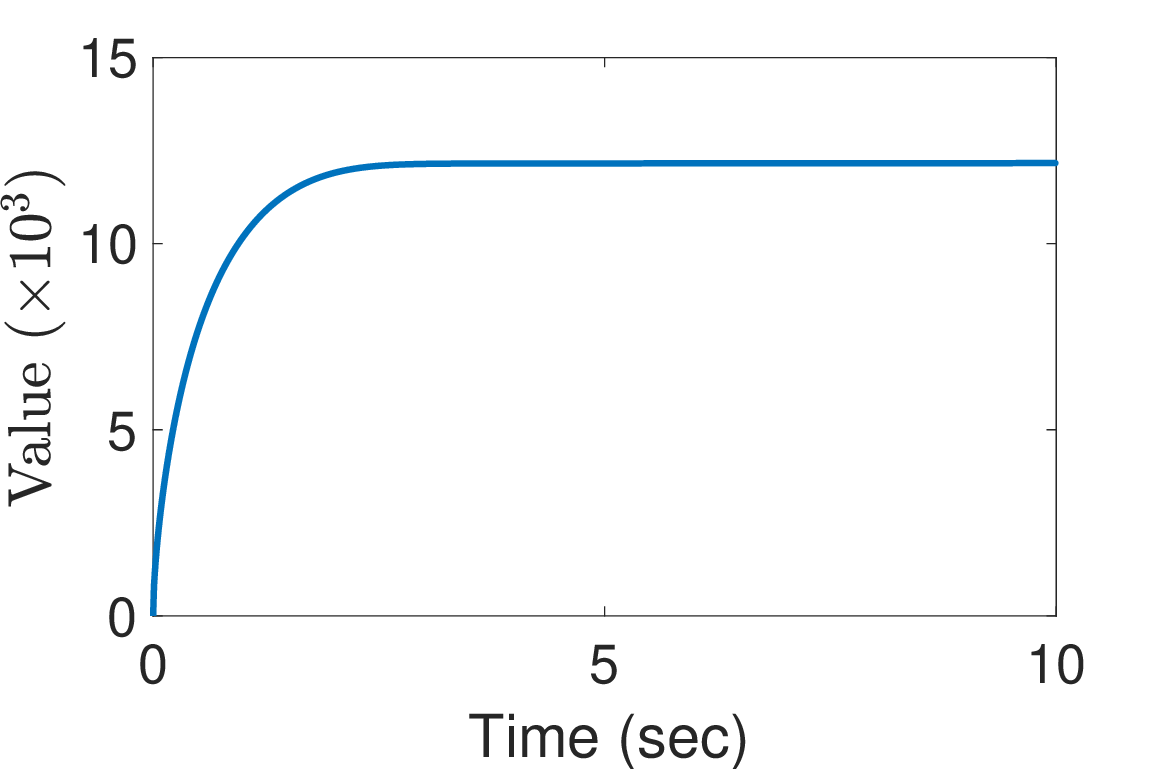}
		  \label{fig:IBVSValue}
	   \end{subfigure}\vspace{-8pt}
	\caption{IBVS: (a) Regulation errors, (b) Control velocities, (c) Value.}
	\label{fig:IBVSControlErrors}
\end{figure*}
\begin{figure*}
        \begin{subfigure}{}
		 \includegraphics[width=0.3\textwidth]{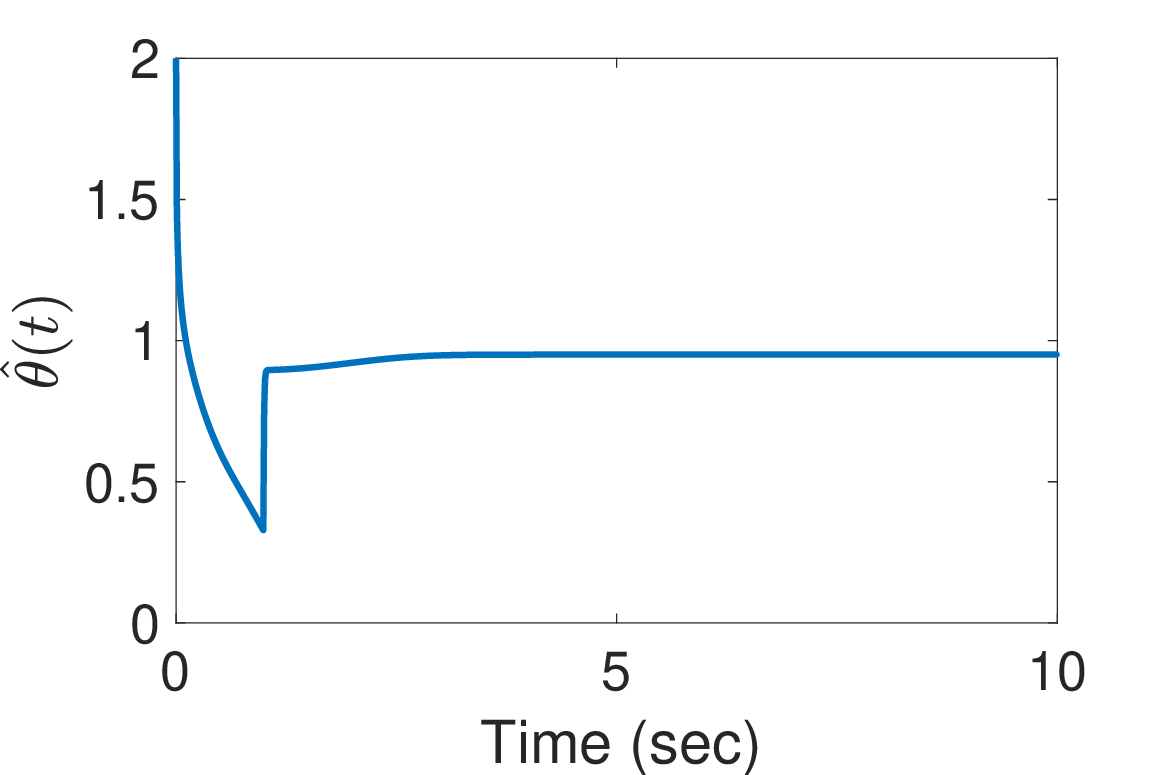}
		 \label{fig:IBVSParameterEstimates}
	      \end{subfigure}
	   \begin{subfigure}{}
		\includegraphics[width=0.3\textwidth]{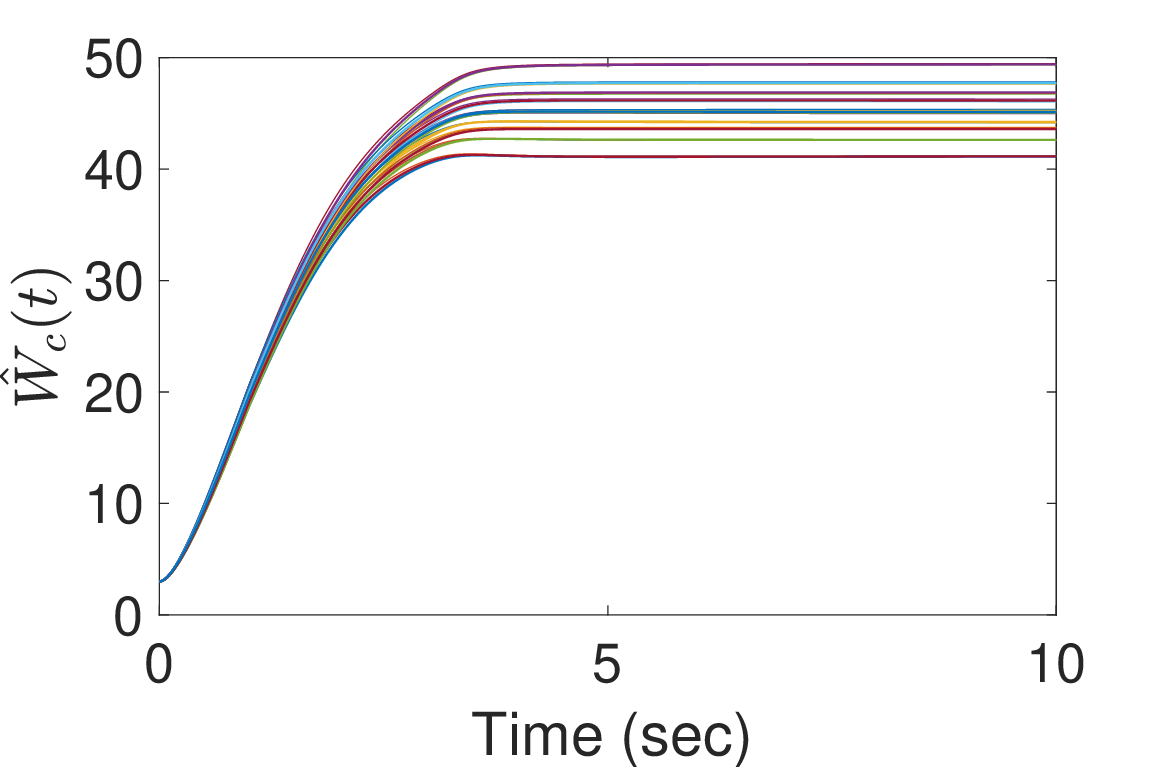}
		\label{fig:IBVSCritic}
	   \end{subfigure}
	   \begin{subfigure}{}
		\includegraphics[width=0.3\textwidth]{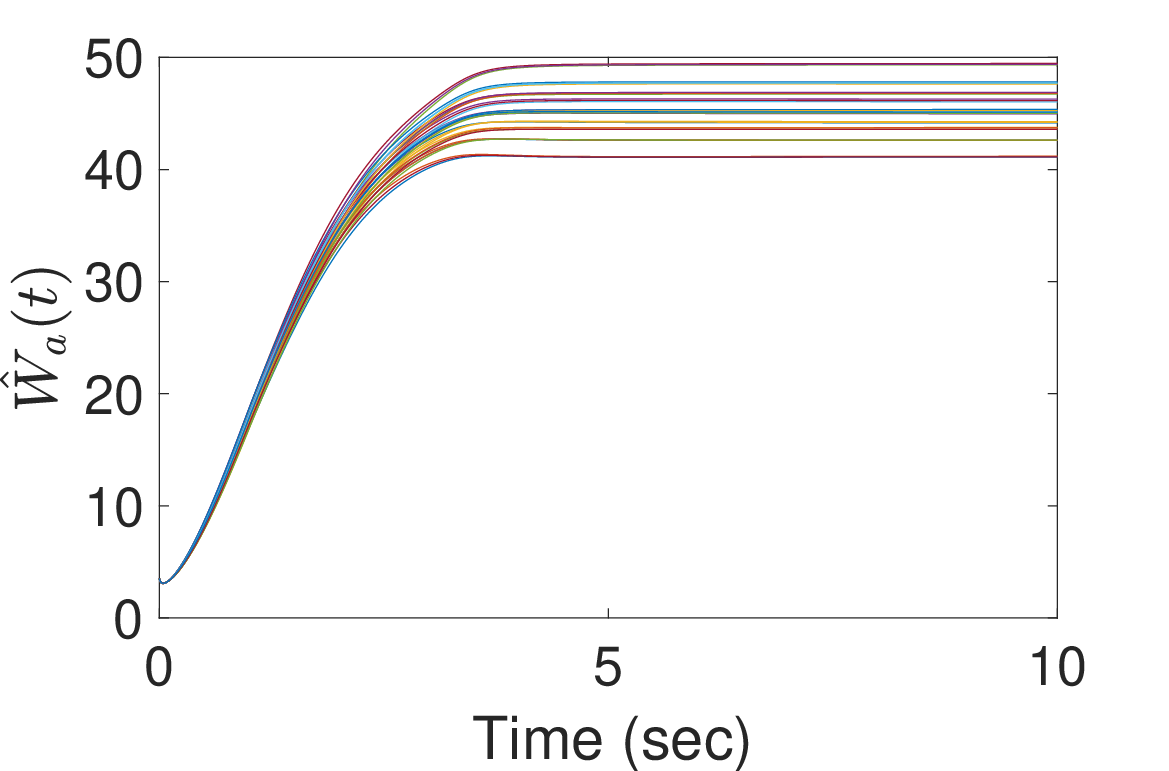}
		\label{fig:IBVSActor}
	    \end{subfigure}\vspace{-8pt}
	\caption{IBVS: (a) Parameter estimates, (b) Critic weights, (c) Actor weights.}
	\label{fig:IBVSEstimates}
\end{figure*}
The dynamics of the IBVS system are given by the system of the form $\dot{x} = g(x,\theta)u$ where $u=[v,\omega]^T$ is the camera velocity consisting of linear velocity $v(t)\in\mathbb{R}^3$ and angular velocity $\omega \in \mathbb{R}^3$. The state $x\in \mathbb{R}^8$ represents the normalized projected coordinates for four points which can be obtained using image pixels and internal camera calibration matrix. The Jacobian matrix $g_i \in \mathbb{R}^{2 \times 6}$ for a feature point is
\begin{equation}
    g_i = \left[\begin{matrix}
        \theta & 0 & -x_1 \theta & -x_1x_2 & 1+x_1^2 & -x_2 \\
        0 & \theta & -x_2 \theta & -1-x_2^2 & x_1x_2 & x_1
    \end{matrix}\right]
\end{equation}
For four feature points the Jacobian matrix is given by $g = [g_1^T \; g_2^T \; g_3^T \; g_4^T]^T \in \mathbb{R}^{8\times 6}$. The parameter $\theta$ is an inverse depth of the feature point which is unknown and varies between $(0,1]$ with time. In the IBVS implementation, $\theta$ is approximated as a constant parameter which is a common practice for IBVS \cite{chaumette2006visual}. The unknown parameter in the system dynamics adds an uncertainty.

For testing the performance of the proposed controller, four points were selected on the image plane with the initial pixel values of $[50 \; 50; 100 \;50; 100\; 100; 50\; 100]^T$ and the desired pixel values of $[825 \; 790; 860 \; 825; 825 \; 860;$ \\$790 \; 825]^T$. The controller parameters are selected as $Q = 800\mathbb{I}_{8\times 8}$ and $R = 60\times \mathrm{blkdiag}\{100\mathbb{I}_{3\times 2},10\mathbb{I}_{3 \times 3}\}$. The basis functions for approximating the value function $V$ is selected to be second order polynomial combinations of elements of $\phi = [\bar{x}_1, \bar{x}_2,..\bar{x}_n]^T$, the initial weights for the critic NN are set to $W_c(t_0) = 3\mathbb{I}_{36 \times 1}$. The parameters of critic NN are found by empirical tuning as $\gamma_c = 2$, $\Gamma = 100$ and $c_1 = 100$. The actor NN weights are initialized to $W_a(t_0) = 3.5\mathbb{I}_{36 \times 1}$. The parameters of actor NN are selected as $\gamma_a = 0.01$ and $\gamma_{a2} = 10.5$. The model parameter update law gains are selected as $\gamma_\theta = 0.1$ and $k_{cl} = 100$. To assure the PE condition of Assumption \ref{ass:assumptionPE}, a probing signal is added to the controller of magnitude $0.01e^{-t}(sin^2(\frac{\pi t}{5})cos(\frac{\pi t}{2}) + sin^2(\frac{2 \pi t}{3})cos(0.1t) + sin^2(-1.2e^{-0.01t})cos(0.5t) + sin^5(e^{-0.1t}))$.

The results of the simulation are shown in Figs. \ref{fig:IBVSControlErrors} and \ref{fig:IBVSEstimates}. The norm of the IBVS regulation errors is shown in Fig. \ref{fig:IBVSControlErrors}(a). The norm of the linear and angular velocities generated by the proposed controller is shown in Fig. \ref{fig:IBVSControlErrors}(b). The control velocities are generated in an optimal manner based on the minimization of the value function whose value is shown in Fig. \ref{fig:IBVSControlErrors}(c). The model parameter weight which is an inverse depth in this case approximated as a constant parameter is shown in Fig. \ref{fig:IBVSEstimates}(a). The parameter is slowly time varying, hence the true model parameter is not exactly identified. The critic and actor NN weights are shown in Figs. \ref{fig:IBVSEstimates}(b)-\ref{fig:IBVSEstimates}(c), which shows that the weights remain bounded, and converge to constant values. Moreover, the actor weights converge to the critic weights. For IBVS controller since the parameter is time-varying but varies between $0$ and $1$, exact parameter estimation may not be achieved, however, the proposed controller drives the regulation error to a small ball around zero in the presence of uncertainties in the image Jacobian matrix.

\subsection{Wheeled Mobile Robot Regulation Controller}
\begin{figure*}
	   \begin{subfigure}{}
		\includegraphics[width=0.3\textwidth]{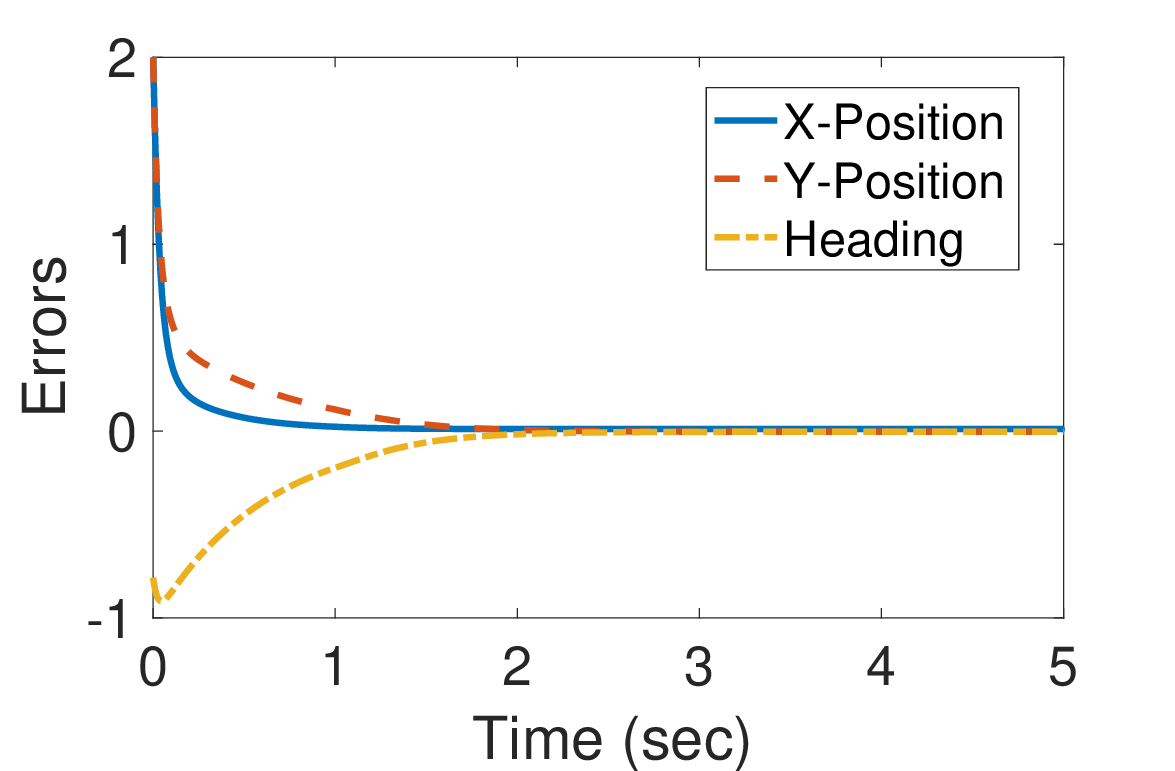}
		\label{fig:WMRErrors}
	   \end{subfigure}
	   \begin{subfigure}{}
		\includegraphics[width=0.3\textwidth]{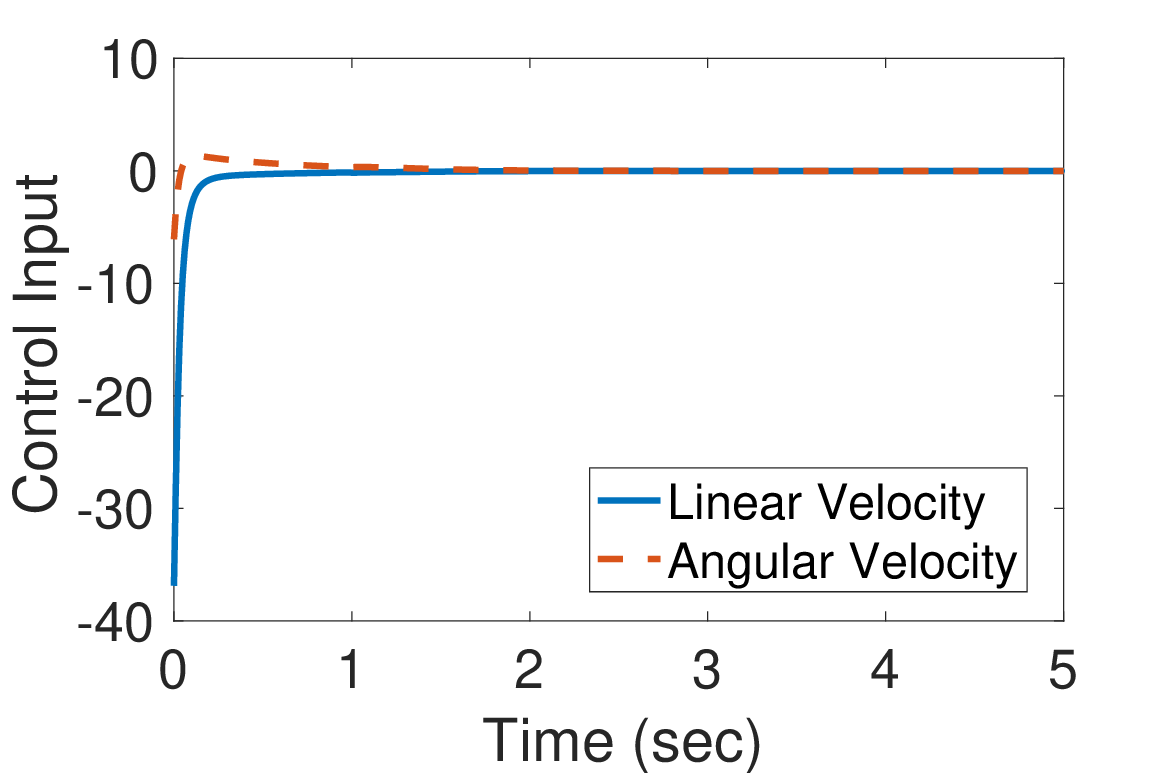}
		\label{fig:WMRControl}
	    \end{subfigure}
	       \begin{subfigure}{}
		  \includegraphics[width=0.3\textwidth]{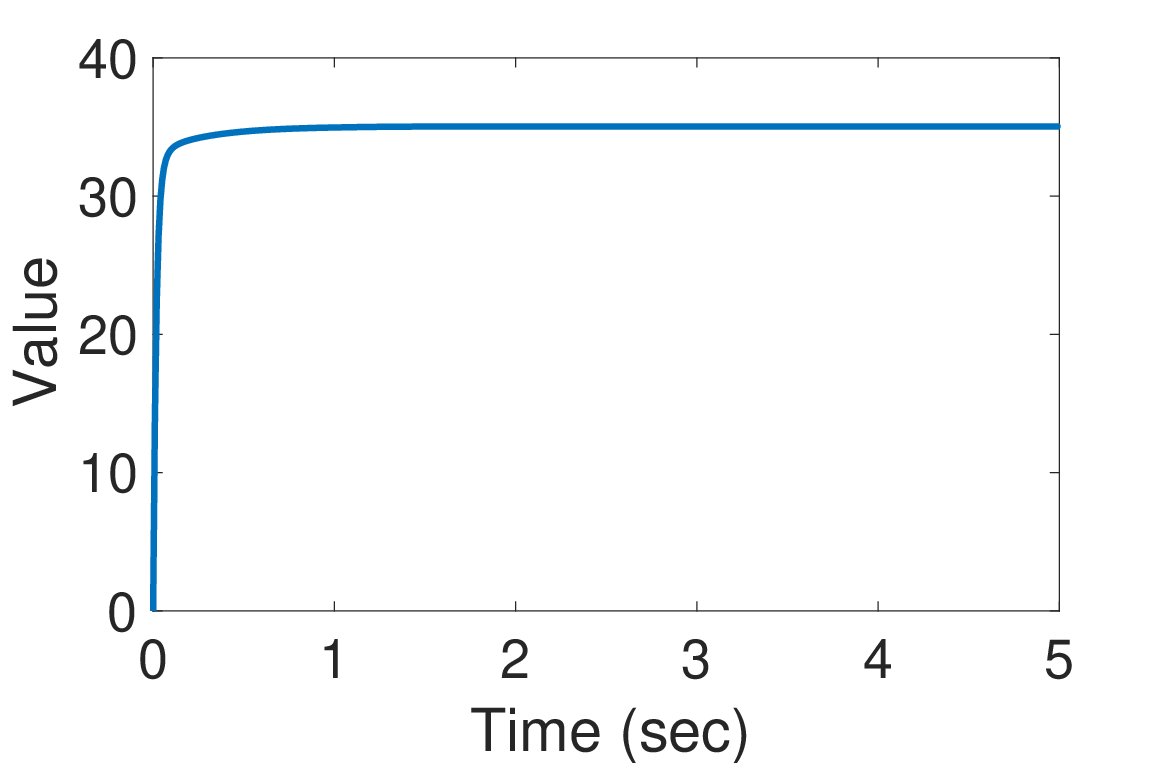}
		  \label{fig:WMRValue}
	       \end{subfigure}\vspace{-8pt}
	\caption{WMR regulation: (a) Regulation errors, (b) Control velocities, (c) Value.}
	\label{fig:WMRControlErrors}
\end{figure*}
The dynamics of the WMR can be written in the following form
%\begin{equation}
$\dot{x} = g(x,\theta)u$,
%\end{equation}
where $x = [X,Y,\psi]^T$ is the 2D position and orientation state, $u = [v,\omega]^T$ are the linear and angular velocities. The Jacobian matrix $g$ can be written as
$g = \left[\begin{smallmatrix}
        acos(\psi) & asin(\psi) & 0 \\
        0 & 0 & b
\end{smallmatrix}\right]^T$, which contains uncertain parameters $a$ and $b$ related to wheel diameter and distance between the wheels \cite{jiang2000robust}. The proposed RL controller is implemented for this dynamics by first formulating the Jacobian in a parametric form as
%\begin{equation}
$\mathrm{vec}(g) = Y\theta$,
%\end{equation}
where $\theta = [a, b]^T$ and 
\begin{equation}
    Y = \left[ \begin{matrix}
        cos(\psi) & sin(\psi) & 0 & 0 & 0 & 0\\
        0 & 0 & 0 & 0 & 0 & 1
    \end{matrix} \right ]^T
\end{equation}
The control task is to regulate the WMR to a desired state of $x_d = [0,0,\frac{\pi}{2}]^T$ from an initial state of $x(0) = [2, 2, \frac{\pi}{4}]^T$. Parameter values of the robot model are selected as $a=1.5$, $b=1$. The proposed adaptive-actor-critic controller is implemented using following parameters: $Q = 2\mathbb{I}_{3\times 3}$, $R = 1\mathbb{I}_{2 \times 2}$. A polynomial basis functions are selected as $\phi = [\bar{x}_1^2,\bar{x}_2^2,\bar{x}_3^2,\bar{x}_1\bar{x}_2,\bar{x}_2\bar{x}_3,\bar{x}_1\bar{x}_3]^T$. The model parameter vector and actor-critic weights are initialized to $\theta(0) = [1,0.5]^T$, $\hat{W}_c(0) = 5\mathbf{1}_{6\times 1}$ and $\hat{W}_a(0) = 10\mathbf{1}_{6 \times 1}$. The controller gain parameters are found by empirical tuning as $\gamma_c = 10$, $\Gamma = 100$, $c_1 = 10$, $\gamma_a = 0.01$ and $\gamma_{a2} = 10.9$. The parameters of the adaptive update law are selected as $\gamma_\theta = 0.01$, $k_{cl} = 50$. A probing signal similar to the previous simulation example is added to the control input.
\begin{figure*}
        \begin{subfigure}{}
		 \includegraphics[width=0.3\textwidth]{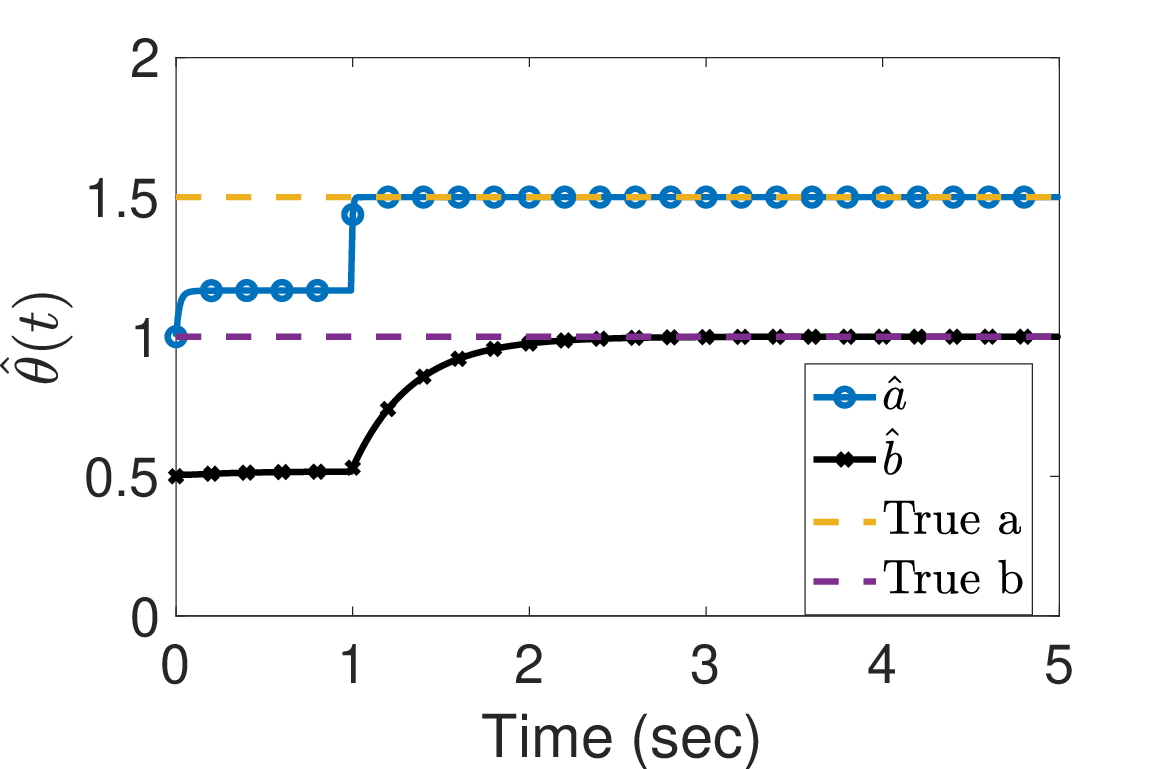}
		 \label{fig:WMRParameterEstimates}
	      \end{subfigure}
	   \begin{subfigure}{}
		\includegraphics[width=0.3\textwidth]{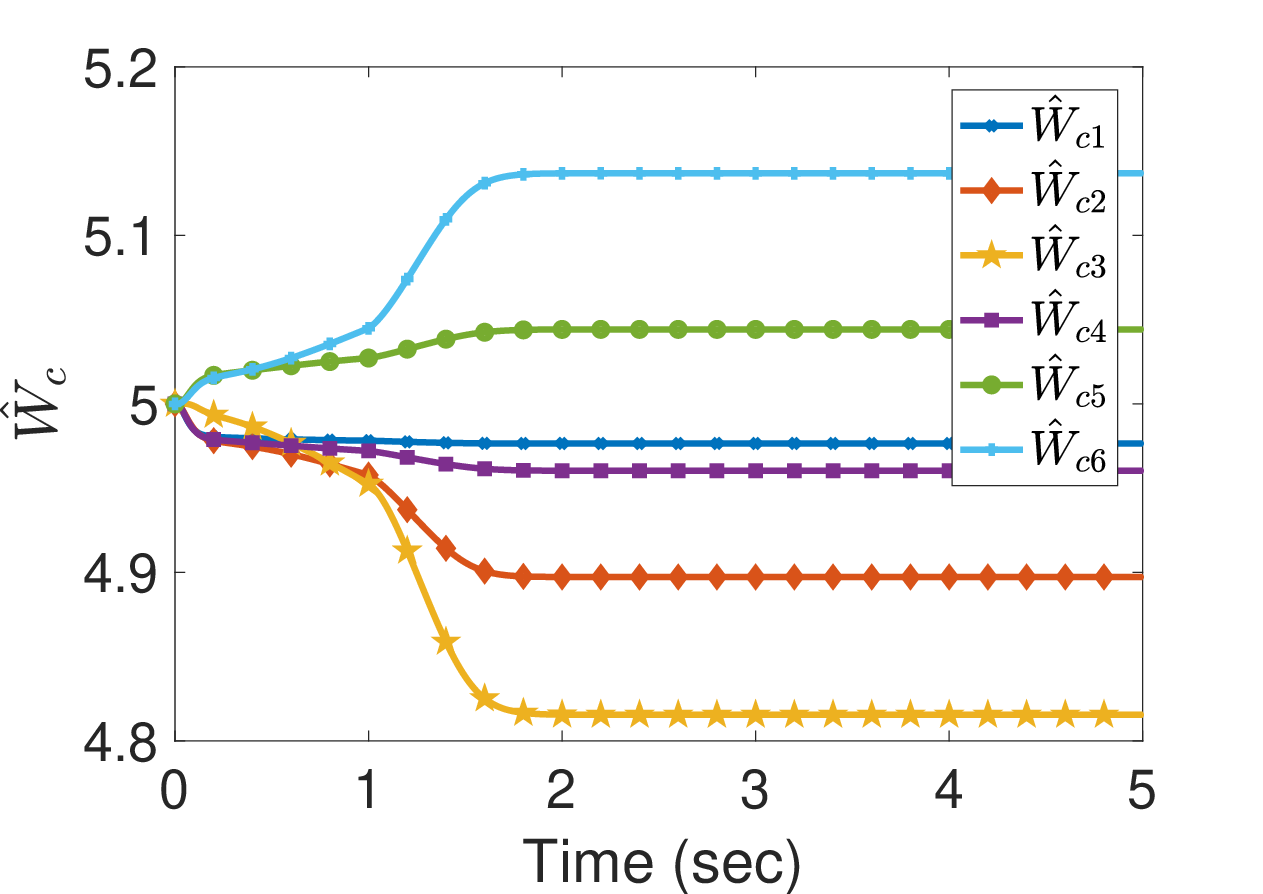}
		\label{fig:WMRCritic}
	   \end{subfigure}
	   \begin{subfigure}{}
		\includegraphics[width=0.3\textwidth]{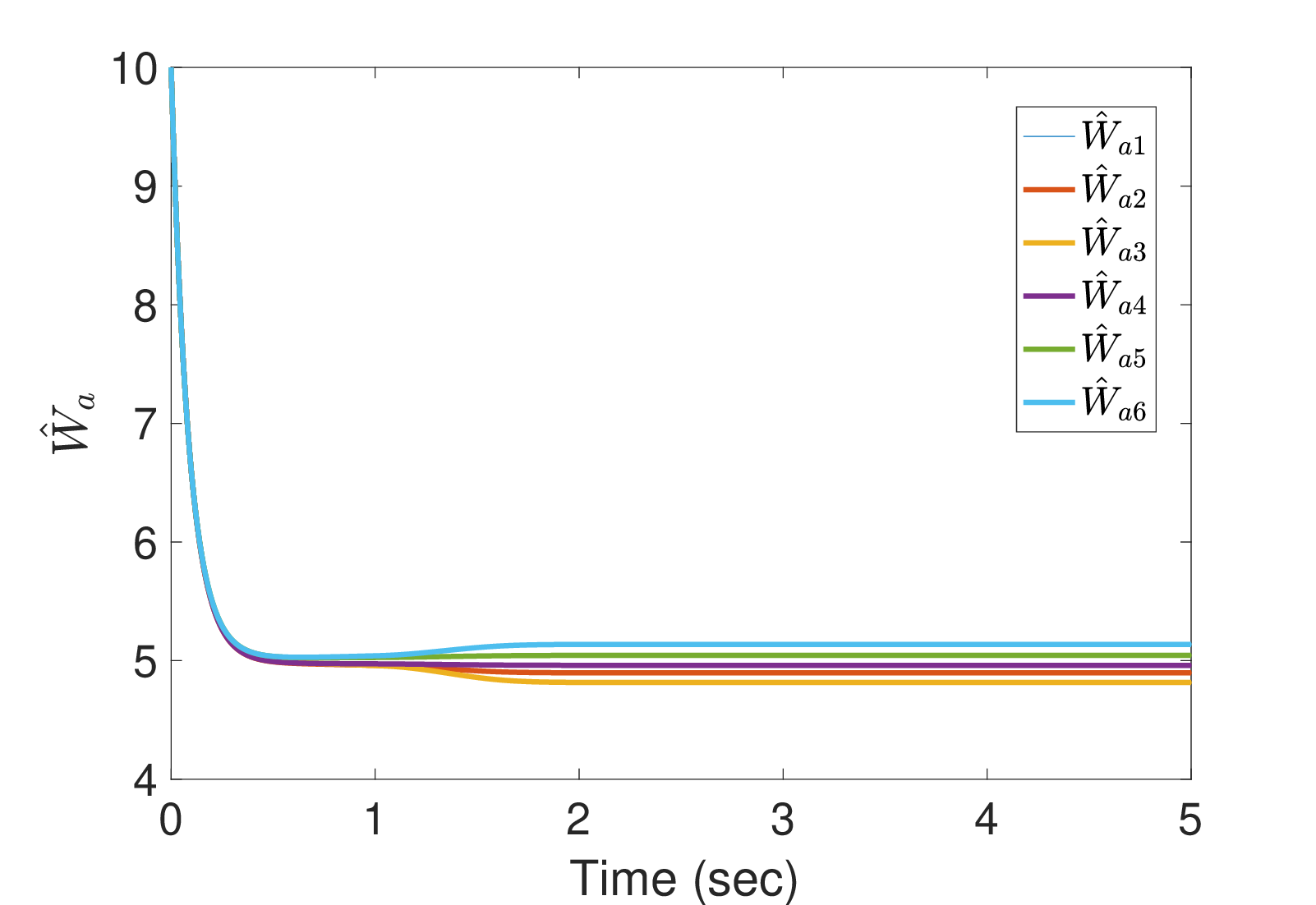}
		\label{fig:WMRActor}
	    \end{subfigure}\vspace{-8pt}
	\caption{WMR regulation: (a) Parameter estimates along with true parameters, (b) Critic weights, (c) Actor weights.}
	\label{fig:WMREstimates}
\end{figure*}
The results are summarized in Figs. \ref{fig:WMRControlErrors}-\ref{fig:WMREstimates}. From Fig. \ref{fig:WMRControlErrors}(a) it is seen that the position and heading angle states are regulated to the desired position and heading angle using linear and angular velocities shown in Fig. \ref{fig:WMRControlErrors}(b) using the proposed RL control policy. The control velocities are bounded and are generated in an optimal manner based on minimization of the value function whose value is shown in Fig. \ref{fig:WMRControlErrors}(c). The WMR model parameters are estimated using CL-based parameter estimation law which achieves the parameter convergence to their true values as seen from Fig. \ref{fig:WMREstimates}(a). The actor and critic NN weights remain bounded and converge to constant values as seen from Figs. \ref{fig:WMREstimates}(b) and \ref{fig:WMREstimates}(c). The actor weights converge to the critic weights.
\section{Conclusion}
In this paper, a reinforcement learning-based policy is developed based on a continuous-time version of the PI architecture for drift-free nonlinear systems with an uncertain $g$ matrix. A concurrent-learning-based adaptive parameter update law is designed to estimate the model parameter and least squared-based update laws are used for actor and critic NN weights. The CL-based parameter update law identifies the parameter using LIP property of the dynamics and history data. Using Lyapunov analysis it is shown that the signals of the closed-loop system are uniformly ultimately bounded. The simulation results on two examples show that the proposed controller can regulate the state to its desired value. In the case of WMR, the constant parameter vector is also identified by the CL-based adaptive update law.
\bibliographystyle{IEEEtran}
\bibliography{RCL_Complete}

\end{document}